\documentclass[runningheads]{llncs}
\usepackage[T1]{fontenc}

\usepackage{graphicx}
\usepackage{svg}
\usepackage{hyperref}
\usepackage{pifont}
\usepackage{multirow}
\begin{document}
\title{Evaluating Pretrained General-Purpose Audio Representations for Music
Genre Classification}
\titlerunning{Evaluating Pretrained Audio Representations for Music Genre Classification}
% If the paper title is too long for the running head, you can set
% an abbreviated paper title here
%
\author{
Kashish Rai\inst{1}\and
Mrinmoy Bhattacharjee\inst{2}
%\orcidID{0000-0002-5348-9378}
}
\authorrunning{Rai and Bhattacharjee}
% First names are abbreviated in the running head.
% If there are more than two authors, 'et al.' is used.
%
\institute{
Dept. of Mathematics, Indian Institute of Technology Patna \and
Dept. of CSE, Indian Institute of Technology Jammu \\
\email{raikashish1002@gmail.com}
}
\maketitle              % typeset the header of the contribution

\begin{abstract}
This study investigates the use of self-supervised learning embeddings, particularly BYOL-A, in conjunction with a deep neural network classifier for Music Genre Classification. Our experiments demonstrate that BYOL-A embeddings outperform other pre-trained models, such as PANNs and VGGish, achieving an accuracy of 81.5\% on the GTZAN dataset and 64.3\% on FMA-Small. The proposed DNN classifier improved performance by 10-16\% over linear classifiers. We explore the effects of contrastive and triplet loss and multitask training with optimized loss weights, achieving the highest accuracy. To address cross-dataset challenges, we combined GTZAN and FMA-Small into a unified 18-class label space for joint training, resulting in slight performance drops on GTZAN but comparable results on FMA-Small. The scripts developed in this work are publicly available.~\footnote{\url{https://github.com/kashishrai12/musicgenre-classification}}

\end{abstract}
\section{Introduction}
\vspace{-0.5\baselineskip}

Music, a universally appealing art form, is often categorized into genres like jazz, pop and others. Music Genre Classification (MGC) helps in efficient music archiving and retrieval. Researchers have used audio feature engineering to create discriminative representations for training classifiers for MGC~\cite{Ru_ICASSP2023,Singh_ESA2022,Pelchat_CJECE2020}. Recent studies show that deep audio foundation models, trained on large datasets via Self-Supervised Learning (SSL), are highly effective at generating discriminating feature embeddings for MGC~\cite{niizumi2023byol,Zhao_ICASSP2022,Spijkervet_ISMIR2021,kong2020panns,hershey2017cnn}. This work explores the impact of carefully designed classifiers using SSL embeddings for MGC.

% Supervised feature learning
Singh et al.~\cite{Singh_ESA2022} compare the effectiveness of standard audio feature representations like Chromagram, Mel-Frequency Cepstral Coefficients (MFCC), and Swaragram~\cite{Singh_FRSM2021} with deep learning models like Convolutional Neural Networks (CNN) and Recurrent Neural Networks (RNN) for the MGC task. Their results indicate that Mel-scale features and Swaragram features are the most generic representations across datasets. Other researchers have also utilized music spectrograms as input features for training classifiers for MGC~\cite{Pelchat_CJECE2020}. Ru et al.~\cite{Ru_ICASSP2023} proposed a multi-modal and multi-label MGC approach by leveraging the inter-genre correlations and the relationship between parallel music and lyrics data.

Recent MGC research has shifted from signal processing-based features to deep SSL embeddings. Zhao et al.~\cite{Zhao_ICASSP2022} introduced the S3T model, trained with a momentum-based contrastive learning approach using the Swin Transformer backbone for MGC and music tagging tasks. Spijkervet et al.~\cite{Spijkervet_ISMIR2021} developed the CLMR framework, based on SimCLR, for music classification from raw audio waveforms, utilizing data augmentation techniques. Niizumi et al.~\cite{niizumi2023byol} proposed BYOL-A, an SSL framework for audio, extending BYOL with audio-specific augmentations and a novel encoder. Kong et al.~~\cite{kong2020panns} introduced Pretrained Audio Neural Networks (PANNs), large-scale pre-trained convolutional networks for audio classification and MGC tasks. Hershey et al.~\cite{hershey2017cnn} proposed the VGGish model, a CNN architecture pre-trained on YouTube data and established a baseline for audio embedding extraction in music information retrieval.

% Motivation and contribution
% \vspace{-0.5\baselineskip}
% Previous works have mostly focused on extracting generalizable SSL embeddings that can be useful for multiple downstream tasks, including MGC~\cite{niizumi2023byol,kong2020panns,hershey2017cnn}. The general practice is to use a simple linear layer to map the SSL embeddings to the target labels of the specific downstream task. To the best of our knowledge, there has been limited focus on designing classifier architectures, training objectives, or learning frameworks to improve the performance obtained using SSL embeddings. In this work, we study the impact of training sophisticated downstream classifiers on the SSL embeddings for improving MGC performance. 
% Rest of the paper is organized as follows. Section~\ref{sec:proposed_work} discusses the methodology employed in this work to design a better performing MGC classifier that uses SSL input embeddings. 
% Section~\ref{sec:experiments_and_results} discusses the different experiments performed and analyzes the results obtained. The paper is concluded in section~\ref{sec:conclusions}.

Previous work has mainly focused on extracting generalizable SSL embeddings , often using a simple linear layer to map embeddings to target labels of the downstream task. However, limited attention has been given to designing classifier architectures or learning frameworks to improve MGC performance using SSL embeddings. This work explores the impact of carefully designed classifiers trained on SSL embeddings to enhance MGC. The paper is structured as follows. Section~\ref{sec:proposed_work} presents the methodology for designing a better MGC classifier with SSL embeddings, Section~\ref{sec:experiments_and_results} covers experiments and results, and Section~\ref{sec:conclusions} concludes the paper.

\begin{figure}[!t]
\centering
\includegraphics[width=\linewidth]{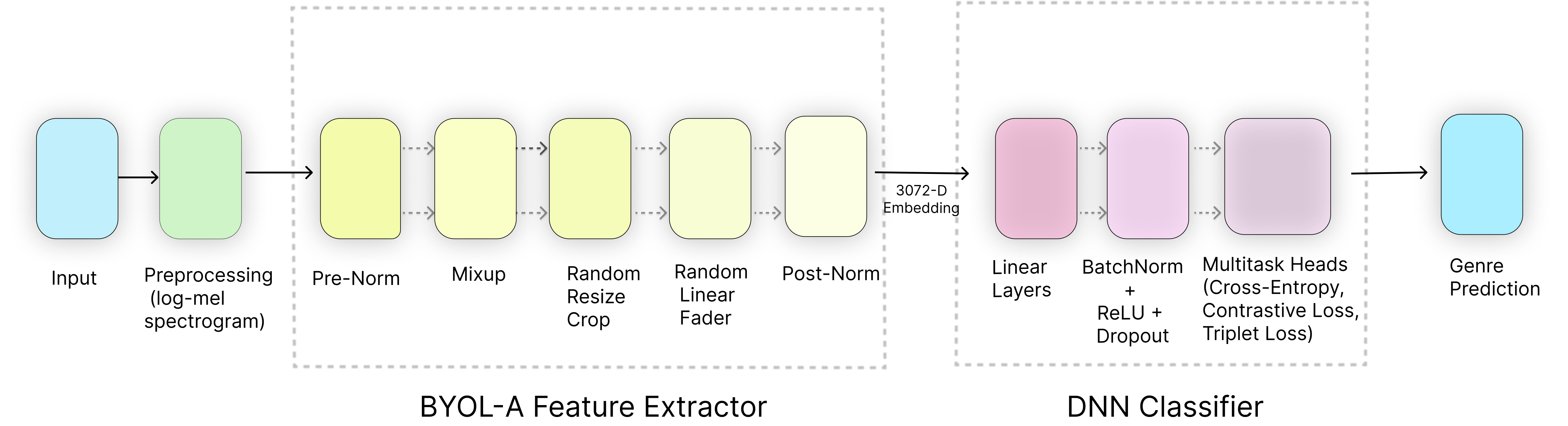}
% \vspace{-0.5\baselineskip}
\caption{Diagram representing the proposed MGC system. The system consists of SSL embedding extractor followed by an DNN classifier for music genre prediction.}
\label{fig:system_diagram}
\vspace{-1.5\baselineskip}
\end{figure}

\vspace{-0.5\baselineskip}
\section{Methodology}
\label{sec:proposed_work}

% Most previous works in MGC have utilized signal processing inspired feature representations like Chroma and Mel-Frequency Cepstral Coefficients~\cite{Singh_ESA2022}. Researchers have also explored self-supervised pretrained models trained specifically for music classification task to obtain informative feature representations for MGC task~\cite{Zhao_ICASSP2022}. 
% We propose a comprehensive two-stage approach: (1) feature extraction using SSL pre-trained models, and (2) classification with a carefully designed Deep Neural Network (DNN) classifier. 
This work uses general-purpose audio SSL models to extract feature embeddings and train carefully designed classifiers to improve MGC performance. 
A block diagram of the proposed system is illustrated in Fig.\ref{fig:system_diagram}. The feature extraction approach used in this work is described below.

\vspace{-0.5\baselineskip}
\subsection{Feature Extraction}
\label{subsec:feature_extraction}

This work compares three different pre-trained SSL models for extracting feature representations, viz., BYOL-A~\cite{niizumi2023byol}, PANNs~\cite{kong2020panns}, and VGGish~\cite{hershey2017cnn}. Each of these models is briefly described next.

% We leverage the Bootstrap Your Own Latent-Audio (BYOL-A) framework proposed by Niizumi et al.~\cite{niizumi2023byol} for feature extraction. 
% In BYOL-A~\cite{niizumi2023byol}, raw audio signals are first transformed into log-mel spectrograms, which serve as the input to a dual-network architecture consisting of an online network and a target network. Each input sample is subjected to a sequence of augmentations, including Mixup (random background blending), Random Resize Crop (simulating pitch/time shifts), and Random Linear Fader (approximating fade-in/out effects). These augmentations ensure the learned representations are invariant to common perturbations in audio data. To capture both fine-grained local features and global context, the BYOL-A encoder combines convolutional feature maps with temporal pooling. Specifically, feature maps are flattened per time frame for local features and passed through a DNN for global feature extraction. These are concatenated and pooled using mean and max statistics over time, resulting in a robust multi-aspect representation. For each $30$s audio track, this process produces a $3072$-dimensional embedding that encodes both temporal and spectral characteristics of the input sound.
In BYOL-A~\cite{niizumi2023byol}, raw audio is converted to log-mel spectrograms and fed into a dual-network architecture with an online and target network. Augmentations like Mixup, Random Resize Crop, and Random Linear Fader are applied to ensure robustness to common audio perturbations. The encoder combines convolutional feature maps and temporal pooling to capture both local and global features. This results in a $3072$-dimensional embedding for each 30-second audio track, encoding both temporal and spectral characteristics.

% \subsection{Details about PANNs and VGGish}

% We utilize the CNN14 model from the PANNs framework proposed by Kong et al.~\cite{kong2020panns} for extracting audio embeddings. PANNs are pretrained on the large-scale AudioSet dataset~\cite{Gemmeke_ICASSP2017} and are designed to capture rich acoustic patterns across diverse sound categories. The input waveform is first converted to a log-mel spectrogram, which is then processed by a deep Convolutional Neural Network (CNN) architecture composed of multiple convolutional blocks and global pooling layers. The model outputs a $2048$-dimensional embedding vector for each input segment, summarizing both spectral and temporal information. These embeddings are extracted in a sliding-window fashion and averaged over time to form a compact representation for each $30$s audio clip.
We use the CNN14 model from the PANNs framework~\cite{kong2020panns} to extract audio embeddings. Pretrained on the large-scale AudioSet dataset~\cite{Gemmeke_ICASSP2017}, PANNs capture rich acoustic patterns. The input waveform is converted to a log-mel spectrogram and processed by a deep CNN with convolutional blocks and global pooling. The model outputs a $2048$-dimensional embedding per segment, summarizing spectral and temporal information. The embeddings are averaged over time to form a compact representation for each $30$s audio clip.

% Hershey et al.~\cite{hershey2017cnn} introduced the VGGish model which is a compact audio feature extractor based on the VGG architecture~\cite{Simonyan_arXiv2014} and pretrained on the AudioSet dataset~\cite{Gemmeke_ICASSP2017}. Input audio is resampled to $16$kHz and converted into log-mel spectrogram patches of $0.96$s duration. Each patch is passed through the VGG-like convolutional network to yield a $128$-dimensional embedding. For our experiments, all patch-level embeddings for a $30$s audio track are averaged to generate a single fixed-length feature vector, providing a lightweight yet informative representation of the input audio suitable for downstream MGC task. In the next subsection, we describe the classifier training details.
Hershey et al.\cite{hershey2017cnn} introduced VGGish, a compact audio feature extractor based on the VGG architecture\cite{Simonyan_arXiv2014} and pretrained on AudioSet~\cite{Gemmeke_ICASSP2017}. Audio is resampled to $16$kHz and converted into $0.96$s log-mel spectrogram patches, which are processed by a VGG-like network to produce 128-dimensional embeddings. For our experiments, embeddings from a $30$s audio track are averaged to create a fixed-length feature vector for MGC tasks. The next subsection discusses the classifier training details.

\vspace{-0.5\baselineskip}
\subsection{Classifier Training}
\label{subsec:classifier_training}

Our approach centers on designing and training a DNN classifier to map SSL embeddings to music genre labels. The architecture is tailored to utilize the high-dimensional feature space and prevent overfitting. The input layer receives the SSL embedding, followed by hidden layers with linear transformations, Batch Normalization, ReLU activation, and Dropout regularization. These layers progressively reduce dimensions, ensuring effective feature abstraction and regularization. The final architecture was determined empirically.

% \subsection{Additional Advances Beyond the Baseline}

% \subsubsection{Alternative Feature Extractors: PANNs and VGGish} 

% To rigorously benchmark the effectiveness of BYOL-A embeddings, we extended our investigation to include alternative self-supervised audio representation models, specifically PANNS and VGGish. 
% For both the GTZAN and FMA-Small datasets, 
% Embeddings were extracted using the SSL models following their prescribed preprocessing and storage protocols. Each set of embeddings was subsequently used as input to the final DNN architecture, ensuring a controlled comparison. This approach allowed us to systematically assess the relative strengths of each these models for MGC.

% \vspace{-0.5\baselineskip}
% \subsection{Loss functions}
\label{subsec:loss_functions}

In the initial experiments for the MGC task, the classifiers are trained using the cross-entropy loss, as defined below.

\vspace{-\baselineskip}
\begin{equation}
    L_{CE} = -\frac{1}{N}\sum_{i=1}^{N} log(p_{i})
\end{equation}
\noindent where, $p_{i}$ is the predicted probability of true class for the $i^{th}$ sample and $N$ is the number of samples in the batch. To further enhance the discriminative capacity of the learned representations, we explored advanced loss functions, viz., contrastive loss and triplet loss. Some previous works have also explored contrastive loss for the classification of music genres~\cite{Ru_ICASSP2023,Costanzi_IWSSIP2024,Spijkervet_ISMIR2021}. The contrastive loss is defined as follows~\cite{Hadsell_CVPR2006}. 

\vspace{-\baselineskip}
\begin{equation}
L_{CL} = \frac{1}{N}\sum_{i=1}^{N} \left\{ (1-y_{i}) \cdot \max\left(m-D_{w}\left(z_{i}^{(1)}, z_{i}^{(2)}\right),0\right)^{2} + y \cdot D_{w}\left(z_{i}^{(1)}, z_{i}^{(2)}\right)^{2} \right\}
\end{equation}

\noindent where, $D_{w}\left(z_{i}^{(1)}, z_{i}^{(2)}\right) = \sqrt{\sum_{k=1}^{n}{\left(z_{i}^{(1)}[k]-z_{i}^{(2)}[k] + \epsilon\right)^{2}}}$ is the Euclidean distance, $\epsilon=1e^{-6}$ is used for numerical stability, $z_{i}^{(1)}$ and $z_{i}^{(2)}$ are the model outputs for the $i^{th}$ input $x_{i}$, $y$ is a binary label indicating similar ($y=0$) or dissimilar ($y=1$) pairs, and $m=1.0$ is the loss margin. We have also explored the Triplet loss, which is described below.

\vspace{-\baselineskip}
\begin{equation}
L_{TL} = D_{w}\left(\mathcal{F}\left(x_{i}^{(a)}\right), \mathcal{F}\left(x_{i}^{(p)}\right)\right)  - D_{w}\left(\mathcal{F}\left(x_{i}^{(a)}\right),\mathcal{F}\left(x_{i}^{(n)}\right)\right) + \epsilon
\end{equation}

\noindent 	where, $\mathcal{F}$ is the model that takes the $i^{th}$ input $x_{i}$ and produces a $d$-dimensional output $\mathcal{F}\left(x_{i}\right)$, and $\epsilon$ is the bias term. Superscripts $(a)$, $(p)$, and $(n)$ denote the anchor, positive, and negative samples, respectively. These losses were incorporated into the training regime as single-task and multi-task setups alongside the standard cross-entropy loss. 
% In the single-task setting, the DNN was fine-tuned to maximize inter-class separation and intra-class compactness in the embedding space. 
The multi-task setup involved multiple output heads, each optimized for a specific loss component, with carefully adjusted weighting parameters. For example, the total loss in the multi-task training, with separate output heads for cross-entropy, contrastive, and triplet losses, is computed as $L = \alpha \cdot L_{CE} + \beta \cdot L_{CL} + \left(1-\alpha-\beta\right) \cdot L_{TL}$. Details of the various configurations explored are provided in subsection~\ref{subsec:contrastive_multitask}.
% These strategies were designed to improve the ability of the model to distinguish between closely related genres, especially in challenging cases with subtle genre boundaries.

% \subsection{Hierarchical Classification}
% \label{subsec:hierarchical_classification}

% A single stage classification of music genres is the predominant approach in the literature. However, we observe that not all genres are distinctly different from each other. Li et al.~\cite{Li_ICASSP2005} have previously demonstrated that using hierarchical taxonomy enhances MGC performance by capturing genre dependencies. Recognizing the potential for genre overlap and similarity, we implemented a hierarchical classification strategy utilizing k-Means clustering. In this approach, the high-dimensional embedding obtained from the pre-trained SSL model was clustered into representative broad genre groups. The classification process was structured in two stages. First, a coarse-level prediction assigned each audio sample to a cluster. Second, a fine-grained cluster-specific classifier distinguished between genres within the assigned cluster. This hierarchical method was explored across a range of cluster configurations to capture broad and nuanced genre relationships. This experiment was designed to address the inherent challenge of separating genres with overlapping acoustic signatures.

\vspace{-0.5\baselineskip}
\section{Experiments and Results}
\label{sec:experiments_and_results}

To evaluate the proposed approach, we conducted experiments to assess the impact of architectural choices, feature extraction methods, and training strategies on MGC performance. Embeddings from BYOL-A~\cite{niizumi2023byol}, PANNs~\cite{kong2020panns}, and VGGish~\cite{hershey2017cnn} were used as input features for the classifier. Two datasets were used for benchmarking: GTZAN~\cite{Tzanetakis_TASP2002}, consisting of $1000$ half-minute music excerpts labeled in $10$ categories, and the FMA Small subset~\cite{Defferrard_fma_dataset2017}, with $8000$ $30$s tracks labeled into $8$ genres. A consistent feature extraction pipeline was applied to both datasets, and embeddings from different models were compared. The experiments and results are detailed in the following subsections.
% These experiments systematically explored the effects of tuning the model 
% % depth, node count, dropout rate, activation functions, batch normalization
% architecture, noise augmentation, alternative embeddings, hierarchical classification, advanced loss functions, and cross-dataset generalization. 

% \subsection{Baseline Systems}
% \label{subsec:baselines}

% \cite{Zhao_ICASSP2022}, \cite{niizumi2023byol}

\begin{figure}[!t]
% First row
\centerline{
\includegraphics[width=0.48\linewidth]{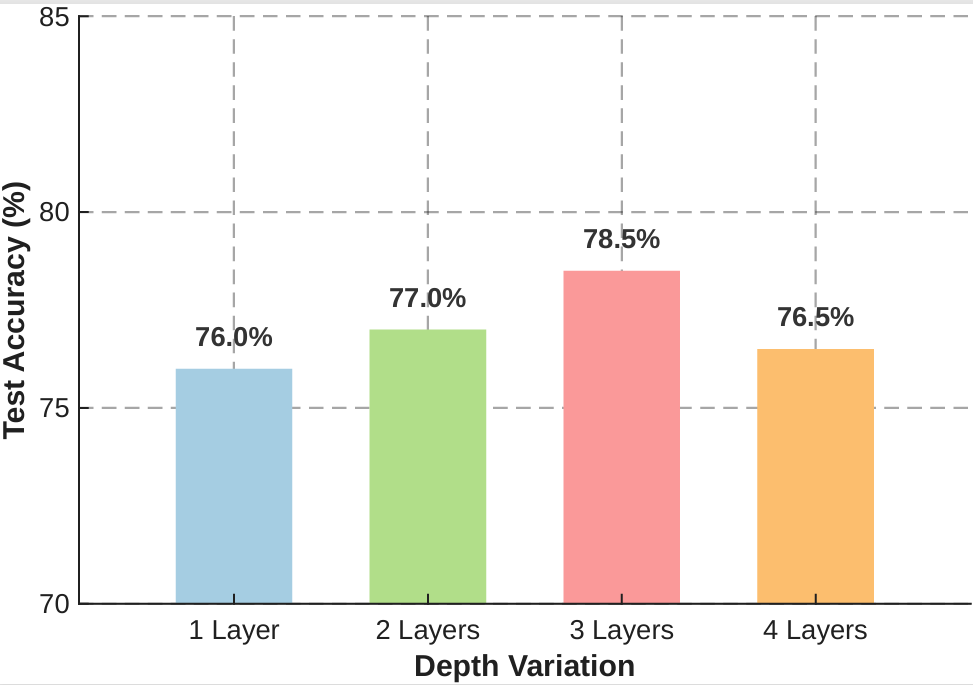}
\hfill
\includegraphics[width=0.48\linewidth]{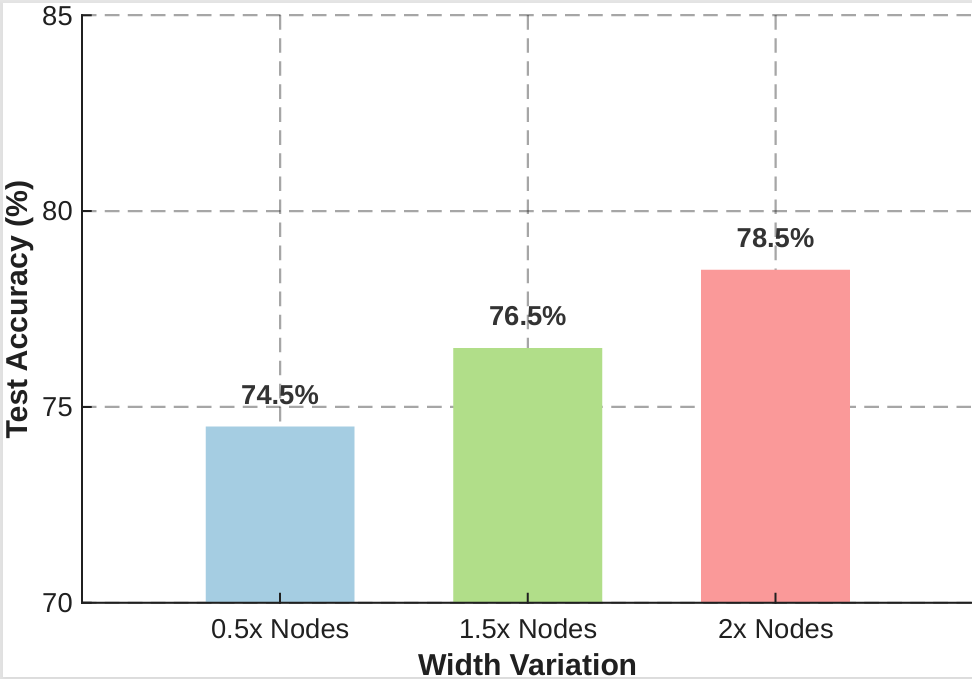}
}
% \centerline{(a) Depth Variation \hspace{4cm} (b) Width Variation}
\centerline{(a) \hspace{5cm} (b) }

% \vspace{0.5cm}
% Second row
\centerline{
\includegraphics[width=0.48\linewidth]{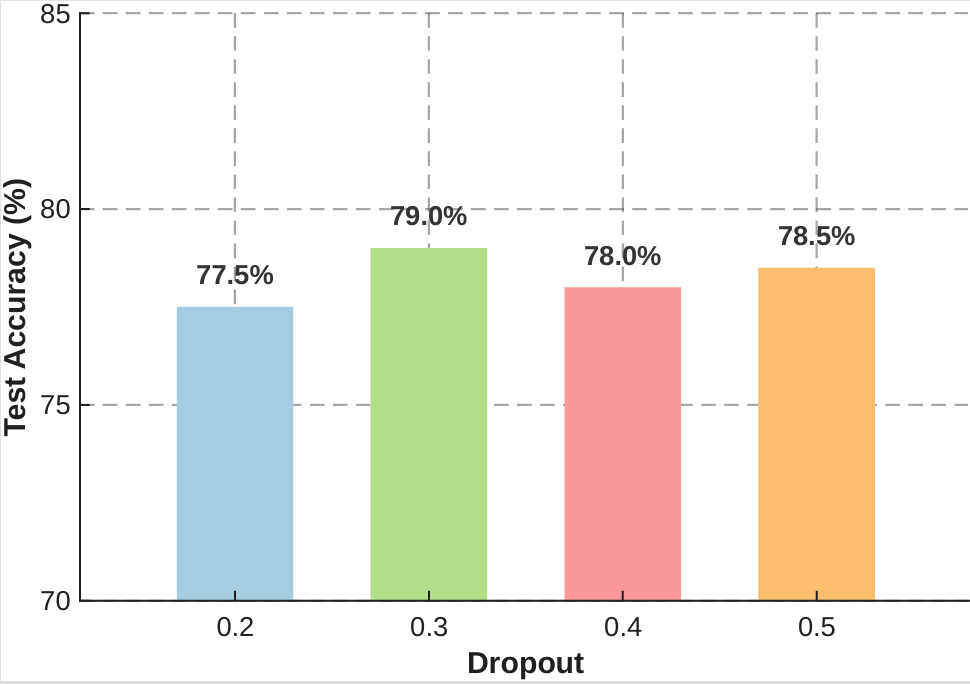}
\hfill
\includegraphics[width=0.48\linewidth]{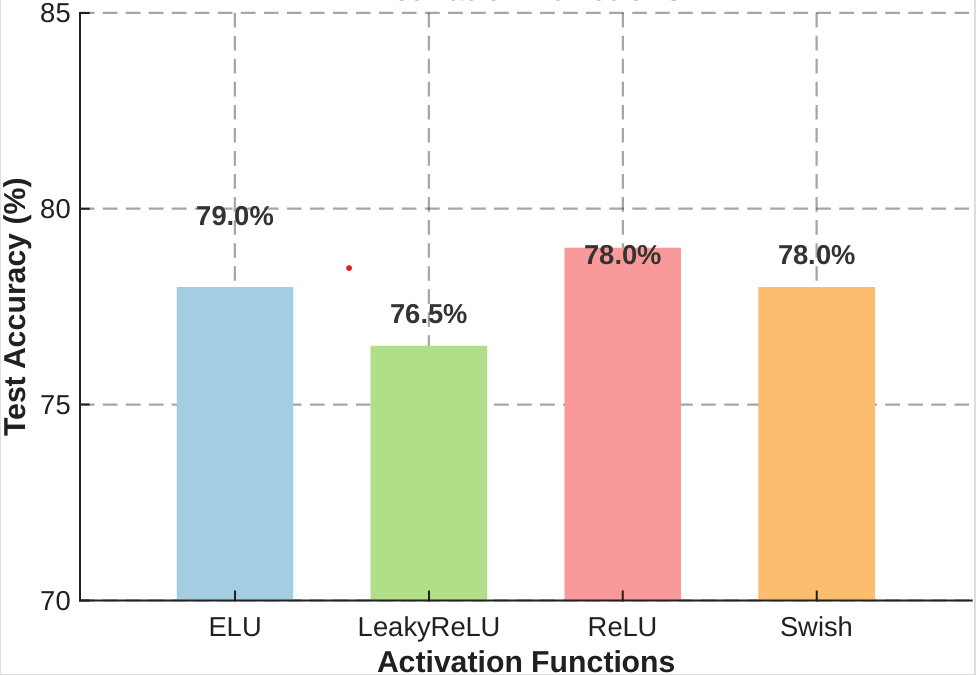}
}
% \centerline{(c) Dropout \hspace{4.3cm} (d) Activation Functions}
\centerline{(c) \hspace{5cm} (d)}

% \vspace{0.5cm}
% Last row (single centered)
\centerline{
\includegraphics[width=0.48\linewidth]{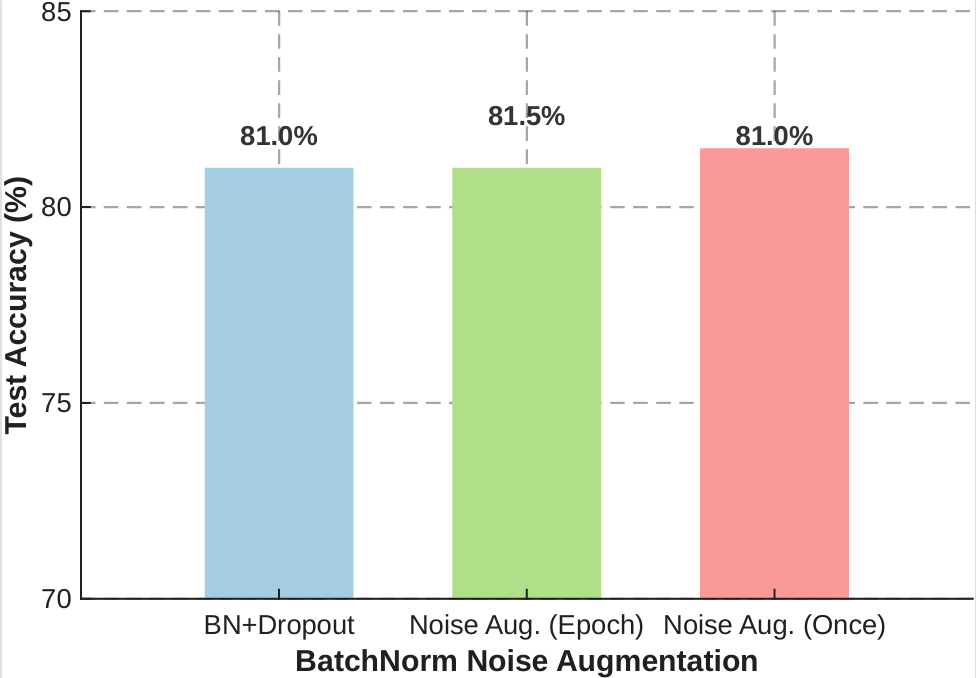}
}
% \centerline{(e) Regularization and Robustness Techniques}
\centerline{(e)}

% \vspace{-0.5\baselineskip}
\caption{Performance of various model architecture tuning experiments performed on the GTZAN dataset using \emph{BYOL-A} embeddings as model input feature. The test accuracy (\%) for different parameter variations is reported.}
\label{fig:architecture_tuning_results}
% \vspace{-2\baselineskip}
\end{figure}

\vspace{-0.5\baselineskip}
\subsection{Selection of DNN architecture}
\label{subsec:model_architecture_parameter_selection}

The impact of hidden layer depth in the DNN classifier was tested with one to four hidden layers (Fig.~\ref{fig:architecture_tuning_results}(a)) using a learning rate of $5e^{-4}$ and batch size of $64$. A three-hidden-layer configuration yielded the best performance, with a test accuracy of $78.5\%$. The effect of node count in this configuration ($128$, $64$, and $32$ nodes) was also explored (Fig.~\ref{fig:architecture_tuning_results}(b)), with doubling the node count achieving the best accuracy of $76.5\%$. No other configurations led to significant improvements.

% A dropout rate of $0.2$ resulted in a test accuracy of $77.5\%$. Higher dropout rates of $0.4$ and $0.5$ also performed well, yielding accuracies of $78\%$ and $78.5\%$, respectively. However, a later experiment with a dropout rate of $0.3$ further improved accuracy to $79\%$. 
% were evaluated for their impact on classification accuracy. 
% In comparison, LeakyReLU and ELU resulted in slightly lower accuracies of approximately $76.5\%$ and $78\%$, respectively. Swish also achieved a test accuracy of $78\%$. 
% These findings indicate that while other activations can perform competitively, ReLU continues to be a robust and reliable choice across experiments.
% Dropout regularization was introduced at varying rates to assess its effect on performance (see Fig~\ref{fig:architecture_tuning_results}(c)). 
% Overall, dropout enhanced generalization, with the best results observed at a rate of $0.3$.
% Activation functions other than Rectified Linear Units (ReLU) were also experimented with, viz., LeakyReLU, Exponential Linear Unit (ELU), and Swish (see Fig.~\ref{fig:architecture_tuning_results}(d)). 
% ReLU achieved strong performance, with a test accuracy of $79\%$. 
% Swish and ELU provided comparable performances, but cannot beat ReLU in overall performance.
% The addition of batch normalization after each linear transformation was found to stabilize training and improve convergence. When combined with a dropout rate of $0.3$, batch normalization helped achieve test accuracies of approximately $81\%$.
Dropout regularization was tested at various rates (Fig.~\ref{fig:architecture_tuning_results}(c)), with the best performance at a rate of $0.3$, enhancing generalization. Different activation functions, viz. LeakyReLU, ELU, and Swish were also evaluated (Fig.~\ref{fig:architecture_tuning_results}(d)), with ReLU yielding the highest accuracy of $79\%$. Swish and ELU performed similarly but did not surpass ReLU. After each linear transformation, applying batch normalization was found to stabilize training and improve convergence, achieving a test accuracy of around $81\%$ when combined with a dropout rate of $0.3$.

\begin{table}[!t]
\scriptsize
\centering
\caption{Test Accuracy (\%) of using different feature extractors on GTZAN and FMA-Small datasets. Results are reported for our best DNN classifier trained on embeddings from each feature extractor.}
\label{tab:feature_extractor_comparisons}
\vspace{-0.5\baselineskip}
{\renewcommand{\arraystretch}{1.3}%
\begin{tabular}{c|c|c}
\hline
\textbf{Feature Extractor} & \textbf{GTZAN}~(\%) & \textbf{FMA-Small}~(\%) \\
\hline
Niizumi et al.~\cite{niizumi2023byol} & $70.1$ & $-$ \\
Defferrard et al.~\cite{Defferrard_fma_dataset2017} & - & $58.0$ \\
\hline
Ours + BYOL-A embeddings & $81.5$ & $64.3$ \\
Ours + PANNs embeddings & $77.0$ & $58.0$ \\
Ours + VGGish embeddings & $79.5$ & $58.04$ \\
\hline
\end{tabular}
}
\vspace{-\baselineskip}
\end{table}

A series of experiments assessed the impact of noise augmentation on model robustness (Fig.~\ref{fig:architecture_tuning_results}(e)). Gaussian noise was added to input embeddings with varied Signal-to-Noise Ratios (SNR) and positions in time. Applying around $20$dB of Gaussian noise at the start of training yielded the best test accuracy of $81.5\%$. Noise augmentation also reduced the number of epochs for convergence from $50$ to about $35$ and helped prevent overfitting.

\vspace{-0.5\baselineskip}
\subsection{Comparison of Feature Extraction Models}
\label{subsec:comparison_feature_extraction_models}

% A comparison was made between BYOL-A embeddings and other pre-trained audio models to evaluate their relative effectiveness. The resuts are reported in Table~\ref{tab:feature_extractor_comparisons}. On the GTZAN dataset, BYOL-A embeddings achieved a test accuracy of $81.5\%$, outperforming PANNS and VGGish, which achieved $77\%$ and $79.5\%$ respectively. BYOL-A embeddings also performed better on the FMA-Small dataset compared to other feature extractors, with a test accuracy of $64.3\%$. 
% Interestingly, the proposed DNN classifier provided approximately $10-16\%$ relative improvement compared to the baseline on GTZAN dataset with all the feature extractors. In the case of the FMA-Small dataset, only the BYOL-A feature extractor could provide approximately $11\%$ relative improvement on the baseline. These results highlight the superior discriminative power of BYOL-A embeddings for MGC compared to PANNs and VGGish embeddings.
A comparison of BYOL-A embeddings with other pre-trained audio models is shown in Table~\ref{tab:feature_extractor_comparisons}. On the GTZAN dataset, BYOL-A achieved $81.5\%$ accuracy, outperforming PANNs ($77\%$) and VGGish ($79.5\%$). BYOL-A also performed better on the FMA-Small dataset with $64.3\%$ accuracy. The proposed DNN classifier provided a relative accuracy improvement of $10-16\%$ on the GTZAN dataset for all feature extractors and $11\%$ on FMA-Small using BYOL-A. These results demonstrate the superior discriminative power of BYOL-A embeddings for MGC and the impact of designing a customized classifier compared to a linear classifier layer.

\begin{table}[!t]
\scriptsize
\centering
\caption{Multitask training results with different loss configurations. Cross-entropy (CE) weights are shown for each classification head, with only one of Contrastive or Triplet used per setup. A \ding{51} indicates the active loss head, with the weight in brackets.}
\label{tab:multitask_final_grouped}
\vspace{-0.5\baselineskip}
{\renewcommand{\arraystretch}{1.3}%
\begin{tabular}{c|c|c|c|c|c|c}
\hline
& \multicolumn{3}{c|}{\textbf{Cross-Entropy ($\alpha$)}} & \textbf{Contrastive} ($\beta$) & \textbf{Triplet} ($\left(1-\alpha-\beta\right)$) & \textbf{Acc~(\%)} \\
\hline
% \multicolumn{6}{c}{\textbf{2 Heads}} \\
% \hline
\multirow{2}{*}{2 Heads} & \ding{51}($0.5$) & - & - & \ding{51} ($0.5$) & - & $79.0$ \\
& \ding{51}($0.5$) & - & - & - & \ding{51} ($0.5$) & $77.0$ \\
\hline
% \multicolumn{6}{c}{\textbf{3 Heads}} \\
% \hline
\multirow{15}{*}{3 Heads} 
& \ding{51}($0.45$) & \ding{51}($0.45$) & - & \ding{51} ($0.1$) & - & $80.0$ \\
& \ding{51}($0.40$) & \ding{51}($0.40$) & - & \ding{51} ($0.2$) & - & $79.5$ \\
& \ding{51}($0.35$) & \ding{51}($0.35$) & - & \textbf{\ding{51} ($0.3$)} & - & $\mathbf{81.5}$ \\
& \ding{51}($0.30$) & \ding{51}($0.30$) & - & \ding{51} ($0.4$) & - & $77.5$ \\
& \ding{51}($0.25$) & \ding{51}($0.25$) & - & \ding{51} ($0.5$) & - & $81.0$ \\
& \ding{51}($0.20$) & \ding{51}($0.20$) & - & \ding{51} ($0.6$) & - & $78.5$ \\
& \ding{51}($0.15$) & \ding{51}($0.15$) & - & \ding{51} ($0.7$) & - & $79.5$ \\
& \ding{51}($0.10$) & \ding{51}($0.10$) & - & \ding{51} ($0.8$) & - & $78.0$ \\
& \ding{51}($0.05$) & \ding{51}($0.05$) & - & \ding{51} ($0.9$) & - & $77.5$ \\
& \ding{51}($0.07$) & \ding{51}($0.63$) & - & \ding{51} ($0.3$) & - & $80.0$ \\
& \ding{51}($0.21$) & \ding{51}($0.49$) & - & \ding{51} ($0.3$) & - & $79.5$ \\
% & \ding{51}($0.35$) & \ding{51}($0.35$) & - & \ding{51} ($0.3$) & - & $79.0$ \\
& \ding{51}($0.49$) & \ding{51}($0.21$) & - & \ding{51} ($0.3$) & - & $78.5$ \\
& \ding{51}($0.63$) & \ding{51}($0.07$) & - & \ding{51} ($0.3$) & - & $81.0$ \\
& \ding{51}($0.35$) & \ding{51}($0.35$) & - & - & \ding{51} ($0.3$) & $80.0$ \\
\hline
% \multicolumn{6}{c}{\textbf{4 Heads}} \\
% \hline
\multirow{1}{*}{4 Heads} & \ding{51}($0.23$) & \ding{51}($0.23$) & \ding{51}($0.23$) & - & \ding{51} ($0.3$) & $79.5$ \\
\hline
\end{tabular}
}
\vspace{-\baselineskip}
\end{table}

% \subsection{Advanced Loss Functions: Contrastive and Triplet Loss}
\vspace{-0.5\baselineskip}
\subsection{Multitask Training}
\label{subsec:contrastive_multitask}

To improve the discriminative power of embeddings, contrastive and triplet loss functions were tested in multitask learning setups with cross-entropy loss. The classifier was extended with multiple output heads, each corresponding to a different loss (e.g., cross-entropy, contrastive, or triplet). Three multitask configurations were explored. First, a $2$-head setup having contrastive or triplet loss plus an auxiliary cross-entropy head for the anchor sample. Second, a $3$-head setup has two auxiliary cross-entropy losses for anchor and non-anchor samples besides the main loss function (contrastive or triplet). Finally, a $4$-head setup was used only for the main loss function as triplet loss with auxiliary cross-entropy heads for anchor, positive, and negative samples. By optimizing a weighted combination of these losses, the model learned representations that were discriminative for genre classification and robust to intra-class variability.

% As can be observed from Table~\ref{tab:multitask_final_grouped}, contrastive loss fine-tuning achieved accuracies up to $79\%$, while triplet loss yielded accuracies between $77\%$ and $80.5\%$, improving with the number of positive and negative samples per anchor. Multitask training with contrastive loss and cross-entropy loss produced the best accuracy of $81.5\%$ when loss weights were optimally tuned. 
% Freezing the embedding layers during fine-tuning significantly reduced accuracy to around $55\%$, highlighting the importance of end-to-end training for optimal performance.
As shown in Table~\ref{tab:multitask_final_grouped}, contrastive loss in a $2$-head setup achieved up to $79\%$ accuracy, while triplet loss ranged from $77\%$ to $80.5\%$. Multitask training with contrastive and cross-entropy loss achieved the best accuracy of $81.5\%$ when loss weights were optimized.

\vspace{-0.5\baselineskip}
\subsection{Domain normalization}
\label{subsec:combined_dataset}

\begin{table}[!t]
\scriptsize
\centering
\caption{Test Accuracy (\%) on GTZAN and FMA-Small using BYOL-A Features}
\label{tab:combined_results}
\vspace{-0.5\baselineskip}
{\renewcommand{\arraystretch}{1.3}%
\begin{tabular}{c|c|c}
\hline
\textbf{Training Dataset} & \textbf{GTZAN}~(\%) & \textbf{FMA-Small}~(\%) \\
\hline
Niizumi et al.~\cite{niizumi2023byol} & $70.1$ & - \\
Defferrard et al.~\cite{Defferrard_fma_dataset2017} & - & $58.0$ \\
\hline

Ours (trained on \emph{GTZAN}) & $\mathbf{81.5}$ & - \\
Ours (trained on \emph{FMA-Small}) & - & $\mathbf{64.3}$ \\
Ours (trained on \emph{GTZAN~+~FMA-Small}) & $78.0$ & $64.25$ \\
\hline
\end{tabular}
}
\vspace{-2\baselineskip}
\end{table}

The GTZAN and FMA-Small datasets have $10$ and $8$ non-overlapping genres, respectively. Such a scenario makes cross-dataset evaluation difficult. We combined both datasets into a unified label space of $18$ classes for joint training to address this. Using the best DNN architecture with BYOL-A feature extraction, the model was trained to predict one of these $18$ classes. As shown in Table~\ref{tab:combined_results}, models trained on the combined dataset performed slightly worse on GTZAN but similarly on FMA-Small. While learning MGC on an augmented label space is more challenging, the BYOL-A embeddings with the designed DNN classifier still performed well. Such results highlight the need for domain adaptation to improve generalization.

\vspace{-0.5\baselineskip}
\section{Conclusion}
\label{sec:conclusions}

This study investigates the effectiveness of SSL embeddings combined with a well-designed DNN classifier for MGC. The approach achieved competitive accuracies on the GTZAN and FMA-Small datasets, with BYOL-A embeddings outperforming PANNs and VGGish in both datasets. Fine-tuning with contrastive and triplet losses provided marginal improvements, while multitask training with contrastive and cross-entropy losses yielded the best accuracy of $81.5\%$. Training on a combined dataset of $18$ genres revealed challenges in domain adaptation. These results emphasize the superior discriminative power of BYOL-A embeddings and the importance of a customized classifier over simple linear layers. Future work will explore embeddings from different SSL model depths and apply domain adaptation techniques to improve cross-dataset performance.

% \begin{thebibliography}{8}

% \bibitem{niizumi2023byol}
% D. Niizumi, D. Takeuchi, Y. Ohishi, N. Harada, and K. Kashino,
% ``BYOL for Audio: Exploring Pre-Trained General-Purpose Audio Representations,''
% \textit{IEEE/ACM Transactions on Audio, Speech, and Language Processing}, vol.~31, pp.~137--150, 2023. DOI: \href{https://doi.org/10.1109/TASLP.2022.3221007}{10.1109/TASLP.2022.3221007}

% \bibitem{kong2020panns}
% Q. Kong, Y. Cao, T. Iqbal, Y. Wang, W. Wang, and M. D. Plumbley,
% ``PANNs: Large-Scale Pretrained Audio Neural Networks for Audio Pattern Recognition,''
% \textit{IEEE/ACM Transactions on Audio, Speech, and Language Processing}, vol.~28, pp.~2880--2894, 2020. DOI: \href{https://doi.org/10.1109/TASLP.2020.3030497}{10.1109/TASLP.2020.3030497}

% \bibitem{hershey2017cnn}
% S. Hershey, S. Chaudhuri, D. P. W. Ellis, J. F. Gemmeke, A. Jansen, R. C. Moore, M. Plakal, D. Platt, R. A. Saurous, B. Seybold, M. Slaney, R. J. Weiss, and K. Wilson,
% ``CNN Architectures for Large-Scale Audio Classification,''
% in \textit{Proc. IEEE International Conference on Acoustics, Speech and Signal Processing (ICASSP)}, pp.~131--135, 2017. DOI: \href{https://doi.org/10.1109/ICASSP.2017.7952132}{10.1109/ICASSP.2017.7952132}

% \end{thebibliography}

% \printbibliography
\bibliographystyle{splncs04}
\bibliography{references}

\end{document}